# Investigating the use of Nanofluids to Improve High Heat Flux Cooling Systems

Thomas R. Barrett[a], S. Robinson[a], K. Flinders[a], A. Sergis[b], Y. Hardalupas[b]

[a]*EURATOM/CCFE Fusion Association, Culham Science Centre, Oxfordshire, OX14 3DB, UK*
[b]*Department of Mechanical Engineering, Imperial College London, London SW7 2AZ, UK*

The thermal performance of high heat flux components in a fusion reactor could be enhanced significantly by the use of nanofluid coolants, suspensions of a liquid with low concentrations of solid nanoparticles. However, before they are considered viable for fusion, the long-term behaviour of nanofluids must be investigated. This paper reports an experiment which is being prepared to provide data on nanofluid stability, settling and erosion in a HyperVapotron device. Procedures are demonstrated for nanofluid synthesis and quality assessment, and the fluid sample analysis methods are described. The end results from this long-running experiment are expected to allow an initial assessment of the suitability of nanofluids as coolants in a fusion reactor.

## 1. Introduction

As engineering design studies progress towards the realisation of commercial energy from nuclear fusion, solutions are being sought which address the extreme thermal management and cooling requirements. Among the most demanding areas of the reactor design are the high heat flux systems, which will receive continuous power densities in the order of 10MW/m$^2$ [1]. The divertor in particular is an area which requires development of high heat flux technology for it to be viable as the reactor power exhaust. To address such requirements, attention must be focussed on coolant specification in combination with heat exchanger design.

The candidate reactor coolants being considered are chiefly helium, water, $CO_2$, molten salts, or dual-cooling using a liquid metal breeder. In developing power plant concepts much attention has been given to helium, including as coolant for the divertor [1], but recently interest has increased in water as it is simpler to implement, makes use of existing power generation technology and has a high power handling capability. The disadvantages of water are a limited power conversion efficiency, high pumping power, and power handling limited by the critical heat flux (CHF, the departure from nucleate boiling with sudden reduction in heat transfer efficiency). However, these limitations could all be mitigated by the use of nanofluids, which are suspensions of 0.001-10% nanoparticles of <100nm size [2]. Water-based nanofluids have the potential to deliver much improved high heat flux cooling [3] while retaining all the advantages of water.

The exciting prospect of nanofluids has motivated this investigation into their suitability as coolants of a fusion reactor. An experiment is being conducted which aims to quantify the long-term effects of a nanofluid when used as working fluid of a HyperVapotron [4] high heat flux device. The experiment sustains a cold nanofluid flow in order to measure nanoparticle stability, settling, and surface erosion. This paper first outlines the potential for adopting nanofluids for high heat flux devices (Section 2), and then describes the experimental rig (Section 3), nanofluid synthesis and characterisation, (Section 4) and the experimental plan (Section 5).

## 2. Potential of Nanofluids for High Heat Flux Cooling in a Fusion Reactor

Nanofluids have been considered before for nuclear reactors [5], although seemingly not for fusion machines. Since the first use of nanofluids in 1993, research into their properties and behaviour has increased rapidly [2], revealing a very promising heat transfer enhancement compared to the base fluid. The statistical literature review in Ref. [3] reports a 10-14% increase in convective/conductive heat transfer and 100-200% increase in the CHF, for the majority of studies. Since there is currently no solution to handle the heat flux at the divertor, it is the dramatic increase in CHF that is potentially the most attractive in fusion engineering. In addition, the enhanced heat transfer efficiency would reduce coolant pumping power and could be used to raise peak cycle temperature and power plant efficiency. However, the mechanisms for this anomalous thermal behaviour are not well understood, and there is thus a need for further underlying research to allow the design of nanofluids for fusion applications.

The safety implications of nanofluid cooling of a fusion reactor are thought to be low. By far the most commonly researched nanofluid is a suspension of alumina in water [3], which is non-toxic but requires some controls as an irritant and respiratory hazard. Aluminium is just outside the low activation threshold, being classifiable as permanent disposal waste after 100 years; however assuming ~1% volume concentrations are used the waste legacy of the nanofluid should be low.

Despite many promising heat transfer studies, the long-term stability of nanofluids, the deposition of particles, and their effect on erosion are not well understood. Considerable research has gone into

---
*Corresponding author email: tom.barrett@ccfe.ac.uk*

methods of stable nanoparticle dispersion [6], and once suitably dispersed they are much less prone to agglomeration than micrometer-sized particles, due to the dominance of electrokinetic effects over van der Waals forces. However, the long-term stability of the dispersion has not been well studied. It is thought that agglomeration of nanoparticles to form larger particles could lead to surface impact, deposition and erosion, although evidence of this in the literature is conflicting [5,7], with very little reported quantifying erosion rate.

Before nanofluids can be considered as an option in the design of a fusion reactor, there is a need to quantify their behaviour and effects in fusion-relevant high heat flux geometries over a prolonged period.

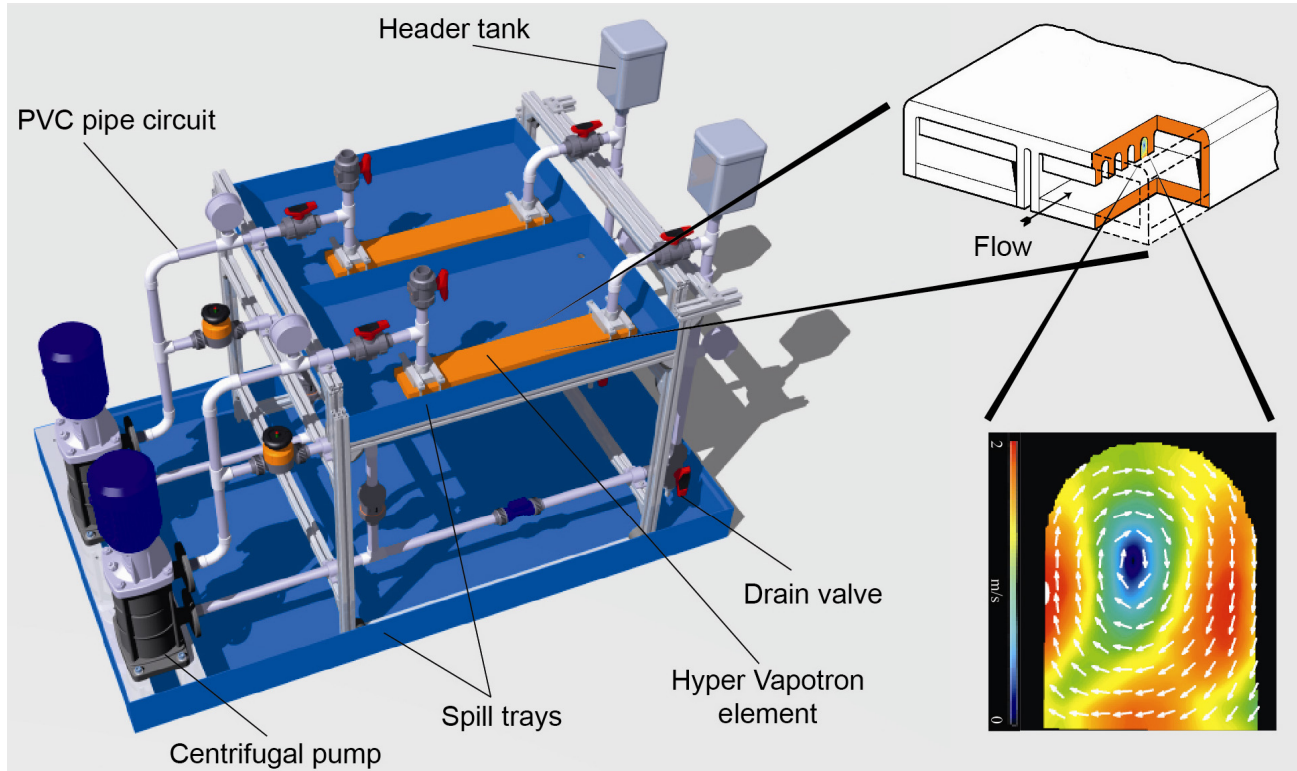

Fig. 1 Nanofluid HyperVapotron experimental rig, with inset HyperVapotron sectional view and experimental results of the velocity field in a single cavity [9].

## 3. Experimental Setup

The principle of the experiment is to recreate the flow in a high heat flux heat exchanger in a laboratory using a nanofluid as the working fluid, under flow conditions that are typical in routine operation of the device but without any applied heating. This cold flow is sustained for a prolonged (indefinite) period and periodic fluid samples are taken to allow monitoring of the fluid content and properties. The nanofluid stability, settling and accumulation are measured as a function of time, and, by comparing with a water-only control case, the effect of the nanofluid on surface erosion rate is quantified.

The experimental apparatus is illustrated in Figure 1. It comprises two identical independent flow loops, one for containing a water-alumina ($Al_2O_3$) nanofluid and the other demineralised water. Each flow loop includes a full-length CuCrZr HyperVapotron high heat flux device [4]. This is an ideal test geometry, as copper-based heat exchangers are known to have suffered from aqueous surface erosion over prolong periods on JET [8]. HyperVapotrons also have a high flow free-stream as well as narrow transverse fins which create relatively stagnant re-circulating flow regions, which could promote particle settling. Figure 1 presents laser velocimetry results showing the typical flow field in a 4mm finned device, part of a study to investigate the HyperVapotron flow field in water compared to nanofluids [9].

In this experiment, two JET Mk1 HyperVapotrons [4] are used, which have 10×48mm channels and 8mm high fins. In order to reduce the flow requirement, only one channel in the device is used, with the other blocked off. The rig is designed to provide a channel flow velocity of 6m/s, which equates to a flow rate of 173litre/min.

The remaining flow loop comprises PVC pipe (including clear sections), header tank and a flow control bypass. A drain valve on the suction side is used for the fluid sampling. A centrifugal pump is used to deliver the required flow at 4.7bar pressure. The work done on the fluid is removed by immersing the HyperVapotrons in water which is passed through a small chiller unit, ensuring that the high flow working fluids are kept at around room temperature. The entire rig is sited in a large spill tray to contain fluids in the event of leaks.

At present, the experimental apparatus is in the phase of manufacture and commissioning. However, in advance of starting the experiment, procedures must be developed for nanofluid synthesis and characterisation.

## 4. Nanofluid Synthesis and Assessment

The planned HyperVapotron experiment uses an $Al_2O_3$-water nanofluid at 1% volume concentration, using <50nm nanoparticles supplied by Sigma-Aldrich Ltd. The two-step method of synthesis is used [6]. First, the nanoparticles are diluted to 1% volume in demineralised water. Second, the fluid is agitated and homogenised using an ultrasonic bath at 38kHz. This sonication step breaks down agglomerates and is therefore key to achieving a uniform and stable suspension.

Following synthesis, the nanofluid is examined using a transmission electron microscope (TEM). The TEM is able to image individual nanoparticles and agglomerates, and is therefore adopted almost universally to assess nanofluid quality. The method used in this work is to place a drop of nanofluid at 1ppm concentration on a carbon film, then dry the fluid in air before imaging using a JOEL 2010 TEM operating at 200kV.

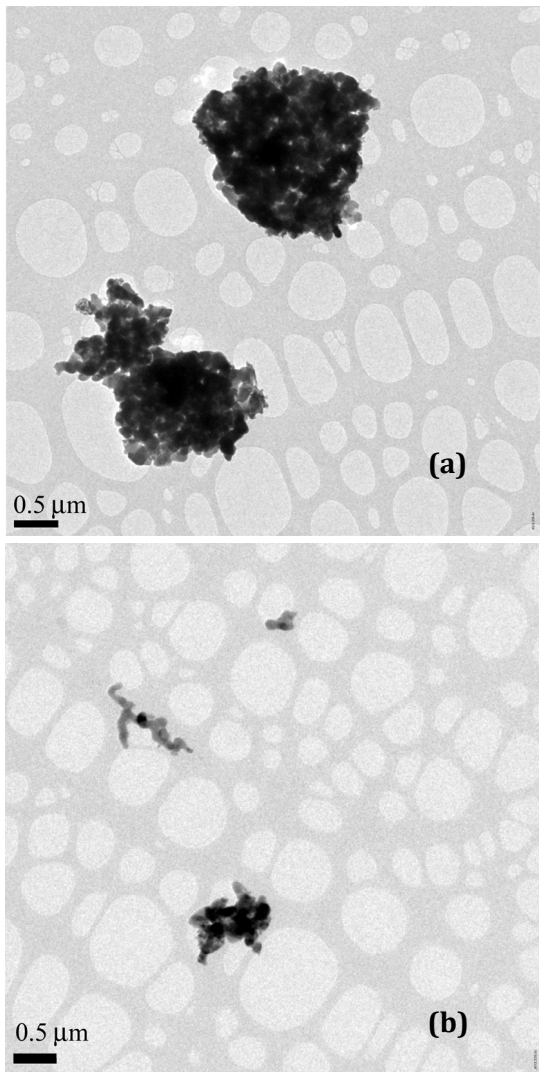

Fig. 2 TEM images at 3k magnification of the $Al_2O_3$ nanofluid after (a) 1hr sonication and (b) 5hr sonication.

Using a combination of zeta potential and TEM analysis, the authors of Ref. [6] concluded that a sonication time of 5 hours produced the most stable dispersion. In order to test this claim, a series of nanofluid samples have been prepared and sonicated for varying lengths of time between 1 and 5 hours. Figure 2 presents TEM images for the 1hr and 5hr samples. The 1hr sample is seen to have large agglomerates, whereas the 5hr sample is generally a much better dispersion. However, even in the 5hr sample, large agglomerates remain; an example is shown in Figure 3 (note the greater magnification compared to Figure 2). If such clustering is ubiquitous, this reinforces the importance of understanding their effects as well as correlating nanoparticle size with heat transfer enhancement.

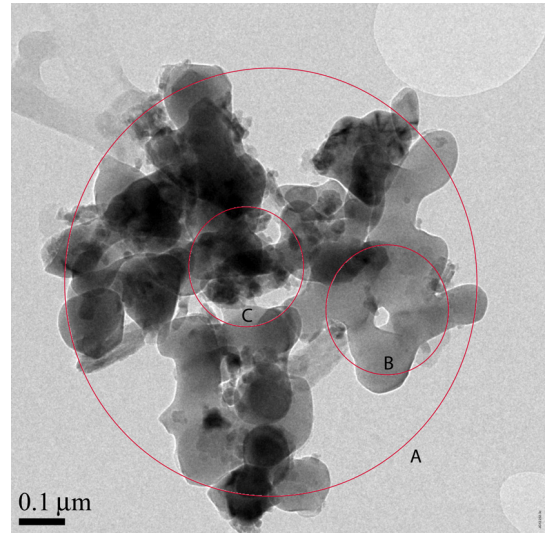

Fig. 3 Large particle agglomerate in the nanofluid sonicated for 5hrs, 15k magnification.

Another important aspect of the TEM is the ability to measure particle material content via the transmitted beam spectrum. This allows a distinction between large $Al_2O_3$ particles, debris or other impurities, and may show whether the $Al_2O_3$ particles are scouring and collecting surface material. In Figure 4, spectral data for the large cluster in Figure 3 shows that the majority is $Al_2O_3$, indicating good purity of the sample and giving confidence in the above method of synthesis.

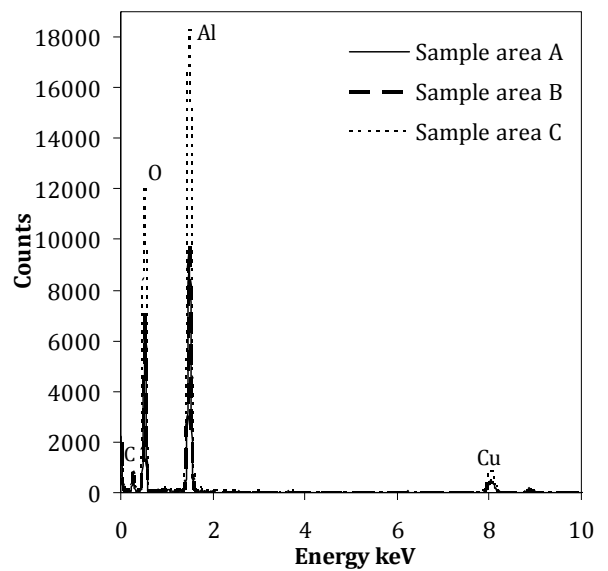

Fig. 4 An example of TEM spectral data, in this case for the particle cluster shown in Figure 3.

## 5. Experiment Plan

As stated above, the experiment in this work is based on taking periodic fluid samples from two sustained HyperVapotron flow loops. On starting the experiment, sampling will initially be relatively frequent to collect data on any short-lived behaviours. Subsequently, samples are expected to be taken monthly. The expected experimental deliverables are:

1. Data as a function of time for
   a) nanoparticle aggregation
   b) particle accumulation or settling
   c) surface erosion of the HyperVapotron or piping
2. Observations of nanofluid long-term behaviour in the flow rig, which may be unanticipated.

The fluid samples are subjected to a number of analysis techniques, as outlined below.

### 5.1 TEM Analysis

The TEM technique described in Section 4 is used to measure nanoparticle size and uniformity and allow an assessment of the rate of particle agglomeration. The spectral data will allow a distinction between nanoparticles, surface debris and other impurities.

For each TEM analysis a 10ml sample will be frozen rapidly in liquid nitrogen to preserve the dispersion. Experimentation performed to date has suggested that freezing and thawing of the nanofluid does not affect the dispersion quality.

### 5.2 Atomic Emission Spectroscopy (AES)

Fluid samples from both flow loops will be analysed for Cu, Cr, C and Al content using inductively coupled plasma AES. The deposition rate of the nanoparticles is assessed by loss of Al from the nanofluid. Erosion rate is quantified by measuring the increase in Cu, Cr or C (hydrocarbons) in the fluid(s). Note, however that this technique will not measure total erosion as it does not account for erosion and re-deposition. For the nanofluid loop, the rate of erosion will be compared with the rate of nanoparticle agglomeration, as it is predicted that increasing agglomeration will accelerate erosion.

### 5.3 Dynamic Light Scattering (DLS)

DLS has the advantage over TEM that it is able to assess the nanoparticle size distribution in the fluid. However, this is based on hydrodynamic size with the assumption of sphericity, which is clearly not always valid (see for example Figure 2 (b)). It also can not distinguish particle clusters from other debris. DLS must therefore be used with care, and is often used in conjunction with a TEM.

Commercially available DLS measurement devices can also typically measure zeta potential, which is often used for assessing nanofluid stability [6]. Thus, the DLS method will be useful in this experiment to provide data on nanofluid quality variation with time.

### 5.4 Internal Inspection

Finally, qualitative observations will be made by visual inspection. An endoscope will be used to periodically inspect the interior of the flow loops, with particular interest on the degree of nanoparticle settling in the narrow HyperVapotron channels. The clear sections of pipe will also be used as sight glasses to observe the deposition of $Al_2O_3$ and/or CuCrZr.

The flow loop pressure drop and flow rate will be carefully monitored as these may indicate changes in internal geometry or surface roughness.

## 6. Conclusion

Nanofluid coolants have the potential to significantly enhance the thermal performance of high heat flux devices in a fusion reactor. No major shortcomings have so far been identified, but there is a need to quantify nanofluid behaviour and effects in fusion-relevant geometries over a prolonged period. An experiment has been prepared which addresses this need. It compares a sustained nanofluid flow in a HyperVapotron with an identical flow loop containing pure water. Procedures have been demonstrated for nanofluid synthesis and quality assessment, and fluid sample analysis methods have been described which will measure the rate of nanoparticle clustering, particle settling and surface erosion. The end results from this long-running experiment are expected to allow an initial assessment of the suitability of nanofluids as coolants in a fusion reactor.

## Acknowledgments

This work was funded partly by the RCUK Energy Programme under grant EP/I501045 and the European Communities under the contract of Association between EURATOM and CCFE. The views and opinions expressed herein do not necessarily reflect those of the European Commission. The authors would also like to extend thanks to the London Centre for Nanotechnology.

## References

[1] P. Norajitra et al., Fus. Eng. Des. **83** (2008), pp893-902.
[2] S. K. Das et al., *Nanofluids: Science and Technology*, First ed., John Wiley & Sons, 2007.
[3] A. Sergis and Y. Hardalupas, Nanoscale Research Letters, **6**(1), 2011, pp391.
[4] H.D. Falter, E. Thompson, Fusion Technol. **29**, 1996, pp584–594.
[5] J. Buongiorno, MNHMT2009-18062, in: Proceedings of the ASME 2nd Micro/Nanoscale Heat & Mass Transfer International Conference, Shanghai, China, 2009.
[6] J. Lee et al., Int. J. Heat and Mass Trans. **51** (2008), pp2651-2656.
[7] C. T. Nguyen et al., *Experimental Investigation of Impinging Jet Heat Transfer and Erosion Effect Using Al2O3-Water Nanofluid*, in: Proceedings of the 6[th] IASME/WSEAS International Conference on Fluid Mechanics, Rhodes, Greece, 2008.
[8] M. Shannon, *Failure of Ion Source Backplates on Jet 2001-2006*, JET Technical Memo, 21[st] December 2006.
[9] A. Sergis et al., *Potential for Improvement in High Heat Flux HyperVapotron Element Performance using Nanofluids*, to appear in: Proceedings of the 24[th] IAEA FEC, San Diego, USA, 2012.